\documentstyle[aps,pra,preprint]{revtex}
\begin{document}
\author{R. E. Cohen}
\address{Carnegie Institution of Washington, 5251 Broad Branch Rd., N.W.,
Washington,%
\\
D.C. 20015 USA}
\title{Periodic Slab LAPW Computations for Ferroelectric BaTiO$_3$}
\date{May 12, 1995}
\maketitle

\begin{abstract}
Linearized augmented plane wave (LAPW) calculations are performed for
periodic (001) and (111) slabs of BaTiO$_3$ to understand the effects of
surfaces on ferroelectric BaTiO$_3$. The (111) slab is found to be much less
stable than the (001) slab. The average surface energies are respectively
3700 erg/cm$^2$ and 1600 erg/cm$^2$. The depolarization field is
sufficiently large in the ideal unrelaxated slab to completely inhibit the
ferroelectric instability. No mid-gap surface states are evident, but there
are surface states in the upper gap in the unrelaxed slab and at the top of
the valence band. The dangling surface Ti bonds self-heal making the Ti-O
surface very reactive and an excellent epitaxial substrate. The charge
density on atoms only one unit cell away from the surface are almost
identical to the bulk.

Keywords: ferroelectric, surface, slab, electronic structure,
depolarization, BaTiO$_3$, thin film
\end{abstract}

\section{Introduction}

Although ferroelectricity is a bulk property, surfaces are crucial for real
ferroelectrics. The termination of an ideal, ionic ferroelectric gives rise
to a huge depolarization field, $E_{d\text{,}}$ of opposite sign to the bulk
polarization, giving a contribution to the energy of $-E_d\cdot P$ large
enough to completely destabilize the bulk ferroelectric state [1] p. 94;
[2]. Properties of ferroelectric films vary with film thickness [3] and
become cubic when thin. Huge particle size effects are observed [4,5] which
again illustrate the importance of surfaces.

Experimental studies of BaTiO$_3$ surfaces are few and many more studies
have focused on SrTiO$_3$ which lacks the complications of ferroelectricity.
Photoemission studies of sputtered and annealed BaTiO$_3$ show evidence of
significant surface defects [6], so intrinsic surface states could not be
easily delineated. Studies of SrTiO$_{3\text{ }}$surfaces show evidence for
minor relaxations by motions of atoms perpendicular to the surfaces [7] and
show no evidence for states in the gap [8]. A tight-binding model for SrTiO$%
_3$ predicted Ti 3d-surface states in the gap [9], inconsistent with the
observations [8], and only after including a coulomb repulsion parameter U$%
_d $ in their model did the states move out of the gap [10]. No
self-consistent calculations for surfaces of BaTiO$_{3\text{ }}$or SrTiO$_3$
have been done. Molecular X$_\alpha $ calculations give some insights into
possible surface effects and give insight into effects of Madelung fields
and charge relaxation at surfaces [11]. Very recent photoemission
spectroscopy on BaTiO$_3$ shows excellent agreement with bulk LAPW
computations of the LDA band structure [12]. See [13] for a review of the
properties of oxide surfaces. First-principles electronic structure
approaches have been very successful for bulk ferroelectrics [14-21] but the
application of first-principles band structure methods to surface properties
is daunting due to the giant computational burden. Here the first set of
accurate electronic structure calculations for periodic BaTiO$_3$ slabs are
presented. It is not possible to fully explore surface relaxations and
interactions between relaxations and bulk ferroelectricity, which is an
ultimate goal of this work. Nevertheless, these results can be considered
benchmark results for faster, more approximate methods, and also give some
insights into the electronic structure of ferroelectric BaTiO$_3$ surfaces.

\section{Method}

Three layer, 15 atom, (001) and (111) periodic slabs of BaTiO$_3$ were
studied using the Linearized Augmented Plane Wave with extra local orbitals
(LAPW+LO) method [22] in the ideal configuration, and the more stable (001)
slab was studied in the ideal configuration, with a tetragonal ferroelectric
distortion, and with some surface relaxations. A vacuum region of the same
thickness as the slab separates the periodic slabs. The ideal slabs are
illustrated in figure 1. The (001) slab has 4mm tetragonal symmetry with
eight-space group operations, and the (111) slab is rhombohedral with 3m
symmetry. The (001) slab has two types of surfaces, both charge-balanced .
One surface contains Ti and O in a ratio of 1:2 and the other consists of
equal amounts of Ba and O. The (111) slab has quite unstable looking
surfaces as shown in figure 1.

The lattice constant (i.e. ideal Ba-Ba distance) was fixed at 7.57 bohr
(4.006 \AA ), which gives an {\it a}-axis length of 7.57 bohr and ${\it c}$%
-axis length of 45.42 bohr (24.035 \AA ) for the (001) slab and 10.70 bohr
for {\it a }and 26.2 bohr for {\it c} for the (111) slab. In order to obtain
accurate energy differences and surface energies, bulk calculations were
also performed in the same symmetry with a tripled supercell with the same
k-points and convergence parameters.

The ferroelectric distortion in the (001) slab used the experimental
tetragonal average displacements. The Ba's were displaced by 0.06 \AA , the
Ti's by -0.1122 \AA\ along {\it c }relative to the O(II)'s, and the O(I)'s
by 0.0288 \AA\ . Displacements were performed in both directions, with Ti
displaced towards the Ba-O surface and towards the Ti-O surface. The
tetragonal phase in bulk BaTiO$_3$ does not consist primarily of
displacements towards the perovskite cube faces, as we are studying here,
but rather displacements towards the cube diagonals, and the tetragonal
structure is a dynamical average with hopping among four [111] directions.
Such disordered configurations are well beyond the possibilities of present
computations for slabs, and must await future faster and probably more
approximate methods. Nevertheless, the general effects of a surface on
ferroelectric properties can be illuminated by the present study.

Two simple surface relaxations in ferroelectric BaTiO$_3$ derived from
experiment [7] on SrTiO$_3$ were also studied. The first was a dimpling of
the Ba-O plane by moving the Ba in by 0.22 \AA . Actually, the ferroelectric
distortion moves this Ba out of the slab away from the O plane by 0.06 \AA ,
so this ``relaxation'' moves the Ba in by 0.16 \AA\ from the O surface
plane. The second ``relaxation'' was to displace the surface Ti out by 0.056
\AA\ as suggested by the RHEED results for SrTiO$_3$. Actually, the
relaxations used are not precisely those determined experimentally since
only the dimpling was included here, and not the change in average layer
spacing. Thus even the limited relaxation effects shown here should be
considered very preliminary estimates.

The muffin tin radii were 2.3 bohr for Ba, 1.75 bohr for Ti, and 1.75 for O,
and the convergence parameter RK$_{\max }$ was set to 7.0. A 4$\times $4$%
\times $2 special k-point mesh was used which gives 3 k-points for the (001)
slab and 10 k-points for the (111) slab. For the (111) slab a 2$\times $2$%
\times $2 mesh was also used and the surface energy differs from the larger
k-point set by only 0.4\%. The matrix order (number of basis functions) for
the slabs was about 2900. These calculations are extremely computationally
intensive. Not only was each iteration in the self-consistent cycle time
consuming (about 11 hours on one processor on the Convex C3 at NCSA) but an
unusually large number of iterations were required to reach self-consistency
due to charge fluctuations across the slab set up by the electric fields
generated across the slab.

\section{Results and Discussion}

Table 1 shows the total energy results. The first order results are the
surface energies. Since we are not studying isolated slabs, but rather
periodic slabs, there are long-range electrostatic interactions between the
periodic images. The results given here have not been corrected for these
interactions, so that the surface energies would be somewhat higher than
given here from this approximation. On the other hand, we have not fully
relaxed any of the slabs, which would lower the surface energy. The numbers
given here can thus be considered only as initial estimates. The energies
shown in Table 1 give the cleavage energy, and half of the cleavage energy
is the average surface energy for the two types of surfaces formed on
cleavage. For the (111) slab an average surface energy of 3700 erg/cm$^2$ is
obtained, and for the more stable (001) surface the energy is 1600 erg/cm$^2$%
. These are quite reasonable values. For comparison, the unrelaxed surface
energy for Cr$_2$O$_{3\text{ }}$is estimated to be 2800 erg/cm$^2$ and 2000
erg/cm$^2$ for the relaxed surface [23]. The surface energy of SiO$_2$ glass
is much lower, about 300 erg/cm$^2$. We have not fully relaxed the surface,
but although the relaxation of the Ba atom in towards the slab reduces the
surface energy by several percent, it is a negligible change considering the
approximations such as neglecting the slab interaction energies.

The depolarization field at the surface does indeed destroy the
ferroelectric instability, and the energy goes up by 6-60 times the bulk
well depth using the average bulk ferroelectric displacements, so the
ferroelectric destabilization energy from the depolarization field for the
perfect slab is many times greater than the bulk ferroelectric instability.

The two surface ``relaxations'' are also quite interesting. The energy gain
on displacing the Ba ions towards the interior of the slab is much greater
than even the bulk ferroelectric well, so that ionic motions on surfaces
could indeed dominate the bulk energetics. This may be the ultimate cause of
the size effects observed for ferroelectrics. Displacing the surface Ti ions
actually increased the energy. This may be due to the ferroelectric
distortion in BaTiO$_3$.

There has been much discussion about possible electronic surface states in
BaTiO$_3$. Calculated band structures for the bulk and ideal slab are shown
in figure 2. No midgap surface states are found, consistent with experiment.
However, the gap is smaller due to the surfaces so unoccupied states at the
top of the gap are predicted. It is possible that these would move up into
the bulk conduction band in a fully relaxed slab. Also, there is an
interesting surface state at M; the topmost occupied state is a pure O p
state on the Ti-O surface, not a Ti d-state as has been discussed from
cluster or approximate calculations [9-11].

The most intriguing results are an examination of the self-consistent charge
densities. Figure 3a shows the differences in total charge density between
the self-consistent LAPW results for the (001) slab and overlapping
spherical ions. The bonds between the O and Ti are clearly evident. Most
interesting is the collapse of the surface Ti bond. Instead of dangling the
charge moves back onto the Ti and the Ti-O surface bonds. This self-healing
leaves the surface highly reactive, and is probably responsible for the
utility of BaTiO$_3$ as a substrate for epitaxial growth and for surface
catalysis. The Ba-O surface shows much less difference from spherical ions,
and is found to be highly ionic. These results suggest that the Ti-O surface
of BaTiO$_3$ is highly reactive due to the possibility of covalent bonding
on the surface, whereas any reactivity Ba-O surface is due entirely to ionic
bonding. The non-bonding O p surface state may also enhance surface
reactivity.

Figure 3b shows the difference between the LAPW (001) slab density and the
self-consistent LAPW charge density for bulk periodic BaTiO$_3$. Large
changes are seen on the surface Ti, and smaller changes on the surface Ba
and O. Most interestingly, the differences the interior Ti and O, which are
only 1 unit cell away from the surface, are almost identical to bulk Ti and
O. This shows that electronic perturbations due the surface are screened
very rapidly in the interior of a crystal. The screening is accomplished
primarily due to polarization and charge redistributions around the surface
Ti, and to a lesser extent, O ions.

\section{Conclusions}

The first precise surface calculations for a ferroelectric are given here
for BaTiO$_3$. These computations are extremely intensive, but will serve as
benchmarks for more approximate but faster methods. It is found that
energetics from surface relaxations sweep the energetics of the
ferroelectric instability in BaTiO$_3$. The (001) surfaces are found to be
much more stable than (111) surfaces, in agreement with experiment. Also in
agreement with experiment no mid-gap surface states are found, but surface
states do appear in the upper gap in the ideal slab and at the top of the
valence band. The surface Ti-bond relaxes back into the surface and
self-heals, making the Ti-O surface as excellent substrate for epitaxial
growth. Even one layer in from the surface the charge density is almost
identical to the bulk indicating that electronic surface effects do not
extend deeply into the solid.

\subsection{Acknowledgments}

Computations were performed on the Convex C3 at NCSA. This research is
supported by ONR research grant N00014-91J-1227.

\section{References}

\begin{enumerate}
\item  Lines M.E. and Glass A.M., {\it Principles and Applications of
Ferroelectrics and Related Materials. } Clarendon Press, Oxford (1977).

\item  Gong Z. and Cohen R.E., {\it Ferroelectrics } {\bf 136,} 113 (1992).

\item  Tsai F. and Cowley J.M., {\it Appl. Phys. Lett. } {\bf 65,} 1906
(1994).

\item  Niepce J.C. {\it Surfaces and Interfaces of Ceramic Materials }
(Edited by L.-C. Dufour), p. 521Kluwer, (1989).

\item  Uchino K., Sadanaga E., and Oonishi K., et al., {\it Ceramic
Transactions } {\bf 8,} 107 (1990).

\item  Cord B. and Courths R., {\it Surface Science } {\bf 152/153,} 1141
(1985).

\item  Hikita T., Hanada T., and Kudo M., {\it Surface Science } {\bf %
287/288,} 377 (1993).

\item  Powell R.A. and Spicer W.E., {\it Phys. Rev. B } {\bf 13,} 2601
(1976).

\item  Wolfram T., Kraut E.A., and Morin F.J., {\it Phys. Rev. B } {\bf 7,}
1677 (1973).

\item  Ellialtioglu S., Wolfram T., and Henrich V.E., {\it Solid State Comm.}
{\bf 27,} 321 (1978).

\item  Tsukada M., Satoko C., and Adachi H., {\it J. Phys. Soc. Jpn. } {\bf %
48,} 200 (1980).

\item  Robey S.W. et al., {\it Ferroelectrics, } {\bf this volume,} (1995).

\item  LaFemina J.P., {\it Critical Reviews in Surface Chemistry } {\bf 3,}
297 (1994).

\item  Cohen R.E., {\it Ferroelectrics } {\bf 150,} 1 (1993).

\item  Cohen R.E., {\it Nature, } {\bf 358,} 136 (1992).

\item  Cohen R.E. and Krakauer H., {\it Ferroelectrics } {\bf 136,} 65
(1992).

\item  Zhong W. and Vanderbilt D., {\it Phys. Rev. Lett. } {\bf 74,} 2587
(1995).

\item  Zhong W., Vanderbilt D., and Rabe K.M. {\it Phys. Rev. Lett. } {\bf %
73,} 1861 (1994).

\item  Zhong W., Kingsmith R.D., and Vanderbilt D. {\it Phys. Rev. Lett. }
{\bf 72,} 3618 (1994).

\item  Yu R., Krakauer H. {\it Phys. Res. Lett.}{\bf ,} in press (1995).

\item  Singh D., {\it Ferroelectrics } {\bf 164,} 143 (1995).

\item  Singh D.J., {\it Planewaves, Pseudopotentials, and the LAPW Method. }
Kluwer Academic Publishers, Boston (1994).

\item  Lawrence P.J. and Parker S.C. {\it Computer Modelling of Fluids
Polymers and Solids } (Edited by C.R.A. Catlow, S.C. Parker, M.P. Allen), p.
219. Kluwer Academic Publishers, Boston (1990).
\end{enumerate}

\newpage\

Table 1. Energetics of BaTiO$_3$ slabs and bulk BaTiO$_3$.

\begin{tabular}{llll}
\hline
& $E+55258Ryd$ & $\Delta E(eV/5atoms)$ & $\Delta E(eV/5atoms)$ \\ \hline
symmetric bulk & $-0.3586$ & $0.0$ &  \\
bulk ferroelectric &  & $-0.005$ &  \\
ideal (001) slab & $-0.1189$ & $1.087$ & $\;0.0$ \\
ferroelectric slab ($+$) & $-0.1119$ & $1.119$ & $\;0.032$ \\
ferroelectric slab ($-$) & $-0.0482$ & $1.408$ & $\;0.321$ \\
ferroelectric slab (Ba in) & $-0.1235$ & $1.066$ & $-0.021$ \\
ferroelectric slab (Ti in) & $-0.1058$ & $1.146$ & $\;0.059$ \\
bulk (111) & $-0.3578$ & $0.0$ &  \\
ideal (111) slab & $+0.5740$ & $4.226$ &  \\ \hline
\end{tabular}

\newpage\

\section{Figure Captions}

1.BaTiO$_3$ slabs studied. Red is Ti, green is Ba, and blue is Ti. a) (111)
slab. b) (001) slab.

2. Band structures for BaTiO$_3$. a) Bulk band structure for cubic BaTiO$_3$
folded into the Brillouin zone for the triple supercell. b) Band structure
for the ideal undistorted three-layer slab.

3. Charge density for BaTiO$_3$ slab. a) Deformation charge density for BaTiO%
$_3$. The image shows the difference in charge density between the
self-consistent LAPW charge density and overlapping PIB ions. Green
represents no change in density (i.e. identical to the ionic crystal).
Partially charged ions, Ba$^{1.63+}$, Ti$^{3.26+}$, O$^{1.63-}$ were used to
generate the ionic charge density. Note the ``healing'' of the surface Ti
dangling bond. It collapses back onto the Ti. b) Difference in charge
density between the ideal slab and bulk BaTiO$_3$. Even though the slab is
only three layers thick, the central layer has almost an identical charge
density to bulk BaTiO$_3$, although there are still some small differences
on the Ba.

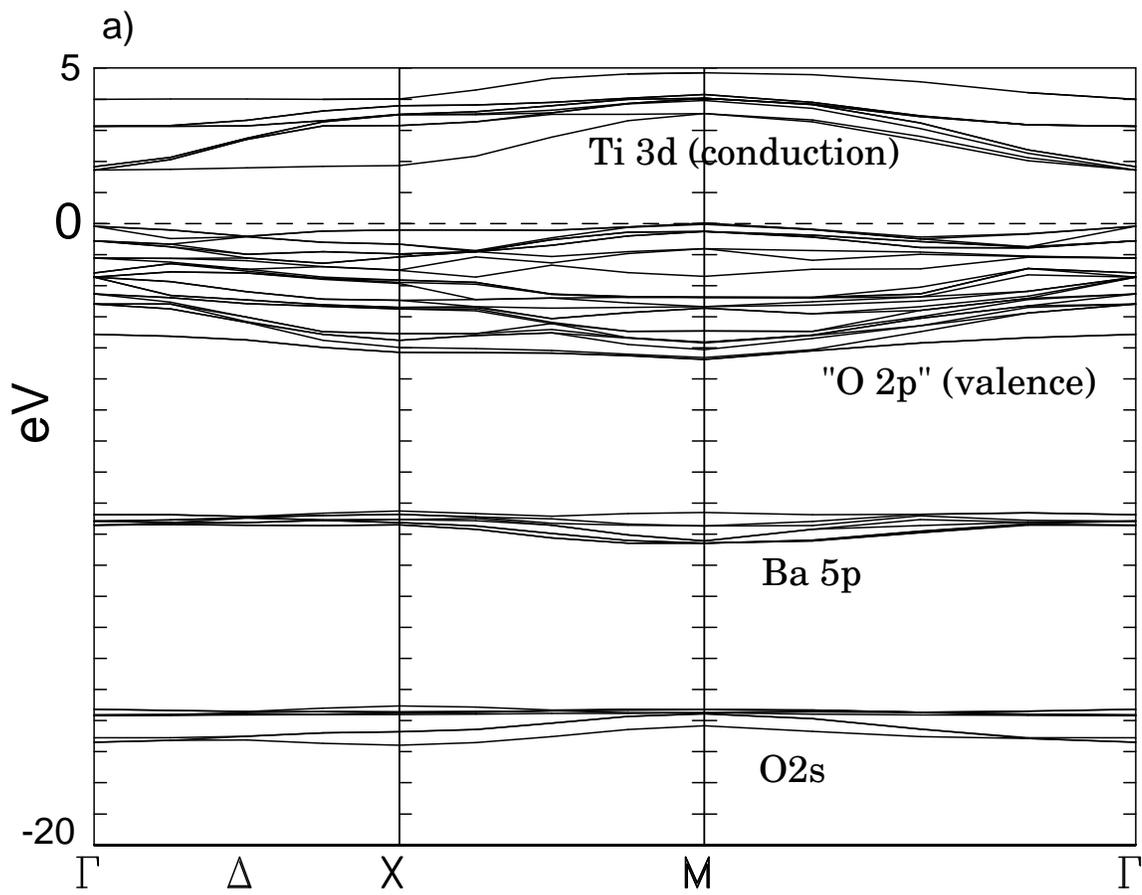
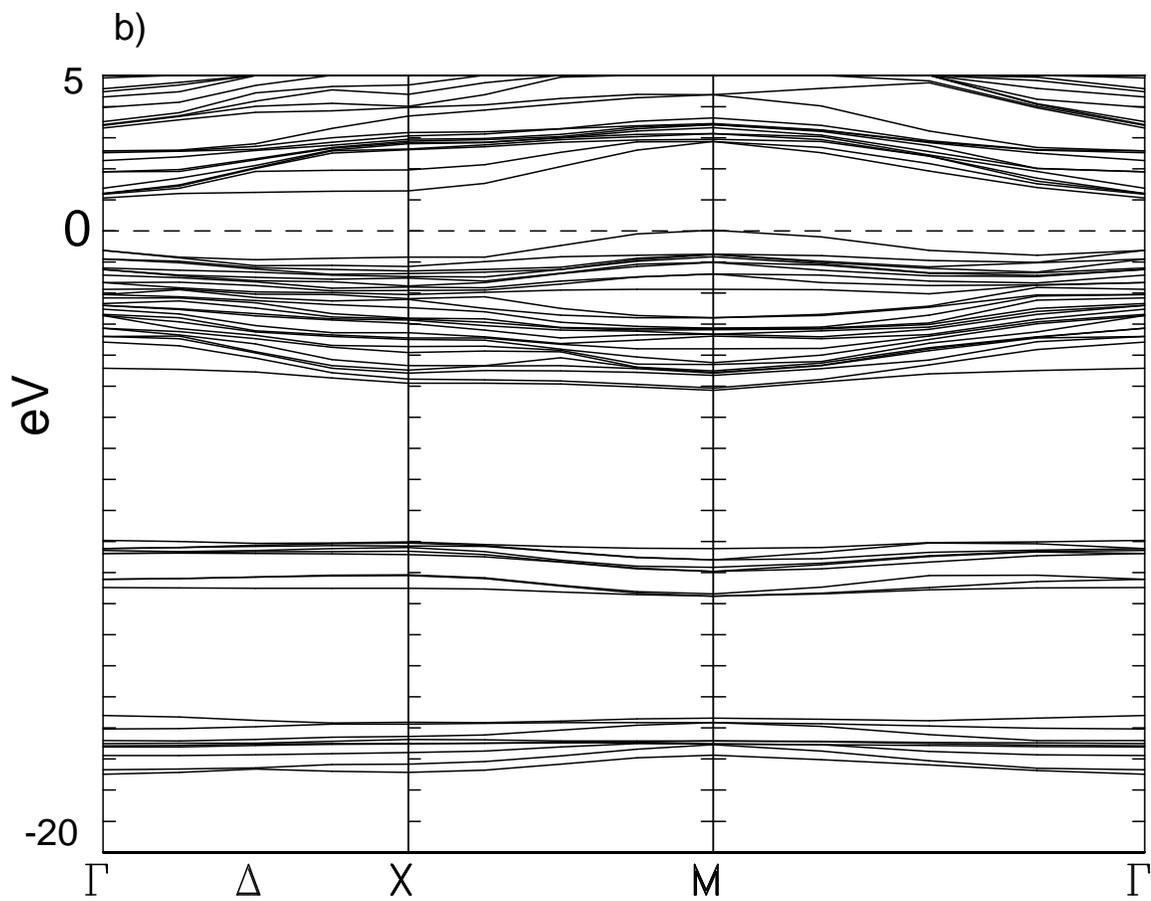
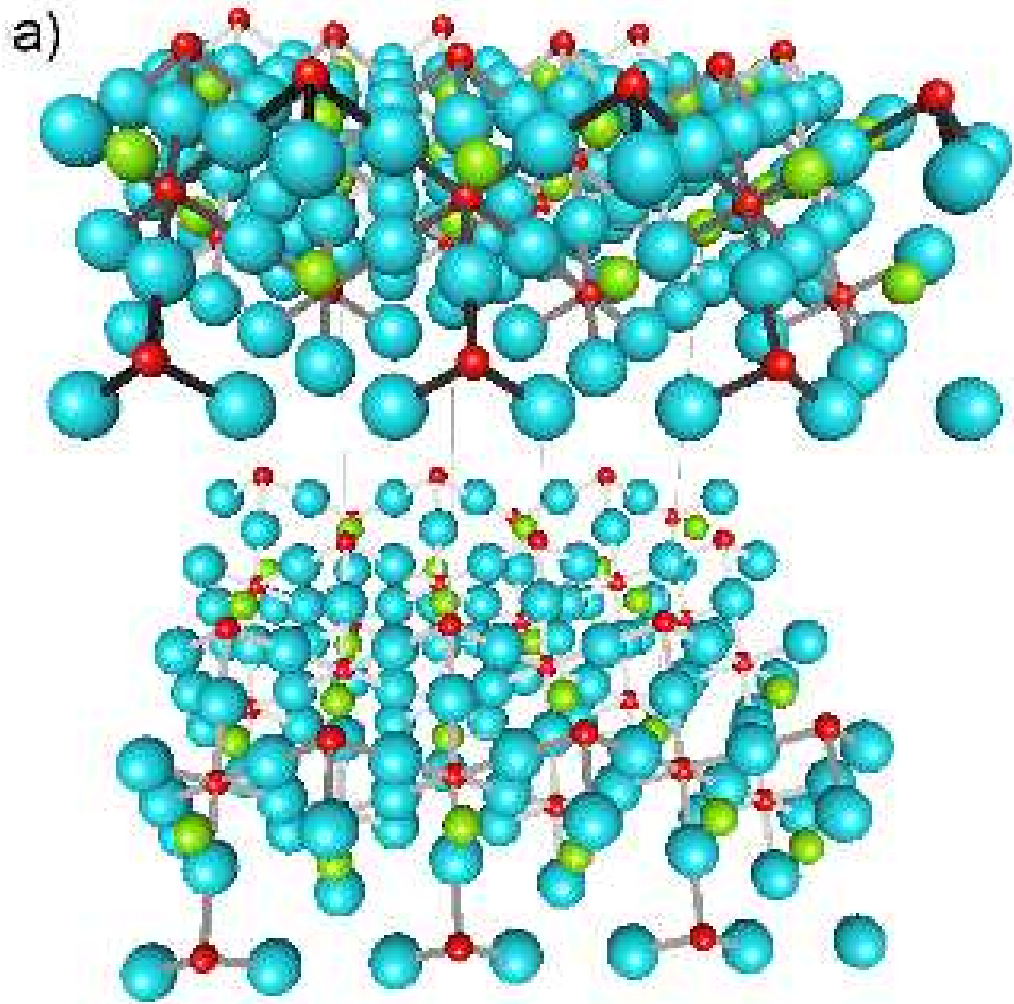
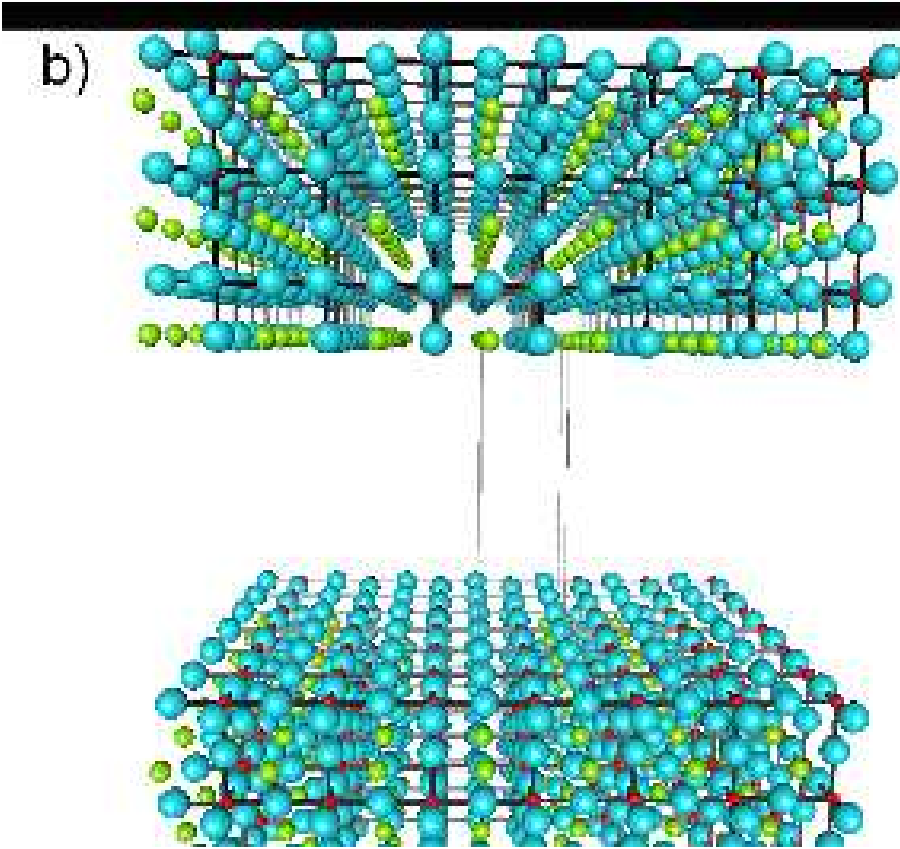
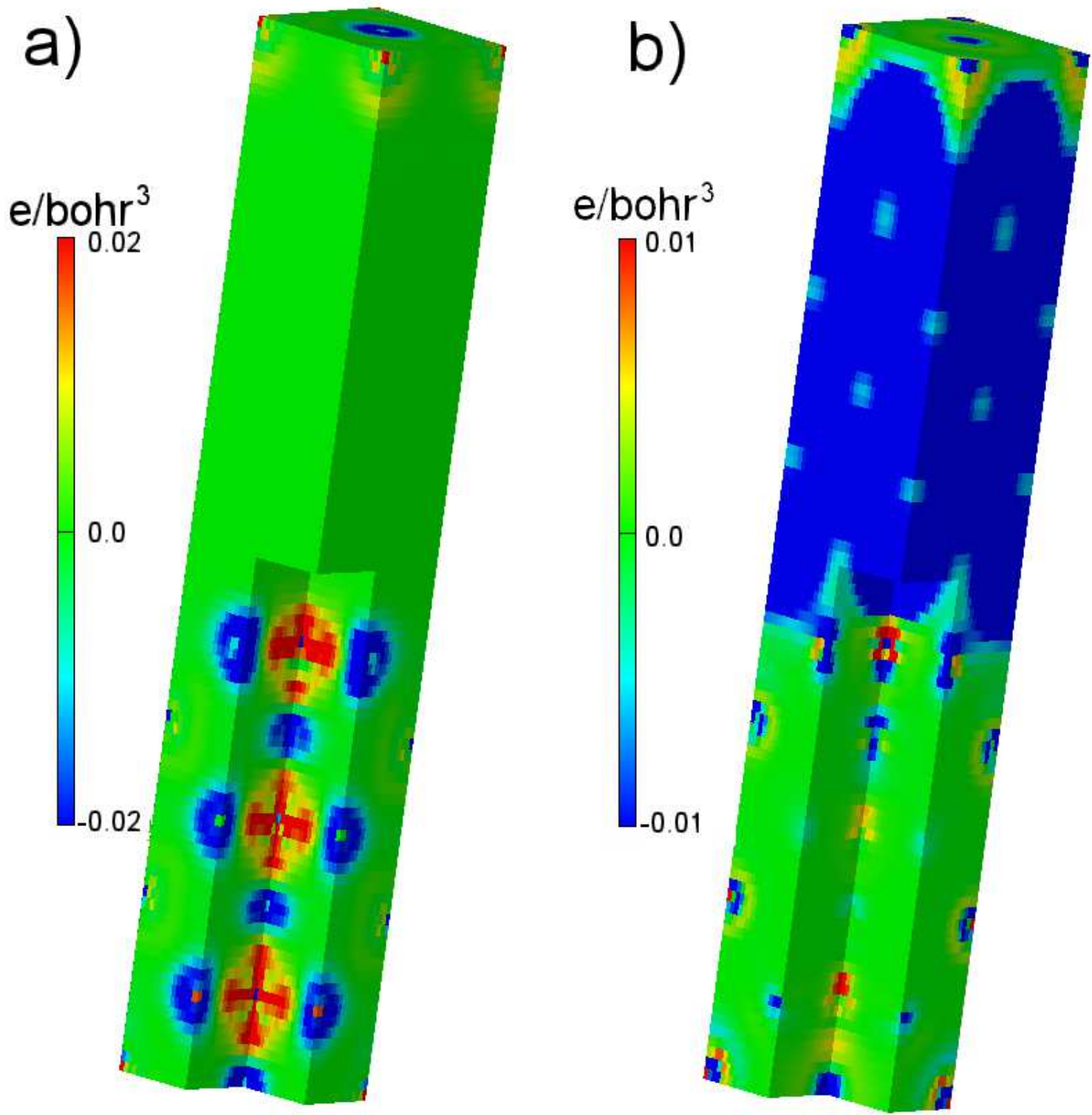
\end{document}